\documentclass[%
 aip,
 amsmath,amssymb,
 reprint,%
]{revtex4-1}

\usepackage{graphicx}
\usepackage{dcolumn}
\usepackage{bm}

\usepackage[utf8]{inputenc}
\usepackage[T1]{fontenc}
\usepackage{mathptmx}
\usepackage{calrsfs}

\usepackage{times}

\usepackage{amsthm}

\usepackage{amsfonts}
\usepackage{amssymb}
\usepackage{graphicx}
\usepackage{bbold}

\usepackage{accents}
\usepackage{graphicx}
\usepackage{gensymb} 
\usepackage{dcolumn}
\usepackage{bm}
\usepackage{amsmath}
\usepackage{graphicx}
\usepackage{yfonts}
\usepackage{float}
\usepackage{mathtools}
\usepackage{comment}
\DeclareMathOperator{\tr}{Tr}

\usepackage[utf8]{inputenc}

\DeclareUnicodeCharacter{2212}{-}

\usepackage{notes2bib}
\bibnotesetup{
note-name = ,
use-sort-key = false
}

\newcommand{\ket}[1]{\left| #1 \right\rangle}

\newcommand{\ketbra}[2]{\left|#1\middle\rangle\middle\langle#2\right|}

\newcommand{\Ket}[1]{{ \vert {#1}  \rangle \!  \rangle}}

\usepackage[normalem]{ulem}
\usepackage{color}

\DeclareMathAlphabet{\pazocal}{OMS}{zplm}{m}{n}

\begin{document}

\preprint{AIP/123-QED}

\title{Experiments on quantum causality}

 \author{K. Goswami}

\email{k.goswami@uq.edu.au}
\author{J. Romero}%
\email{m.romero@uq.edu.au}
\affiliation{ 
 Australian Reserch Council Centre of Excellence for Engineered Quantum Systems, School of Mathematics and Physics,\\ University of Queensland, QLD 4072 Australia
}%

\date{\today}

\begin{abstract}
Quantum causality extends the conventional notion of fixed causal structure by allowing channels and operations to act in an indefinite causal order. The importance of such an indefinite causal order ranges from the foundational---e.g. towards a theory of quantum gravity---to the applied---e.g. for advantages in communication and computation. In this review, we will walk through the basic theory of indefinite causal order and focus on experiments that rely on a physically realisable indefinite causal ordered process---the quantum switch.

\end{abstract}

\maketitle

\section{\label{sec:level1}Introduction}
Causality is central to the way that we model the world. There is a definite relationship between cause and effect—effects depend on their causes. This relationship between cause and effect is called causal order, and this is established by observing changes in the effect given some  freely chosen intervention on the cause. The effects and interventions happen in the physical stage that is space-time \cite{milburn2018classical}.  In conventional quantum physics, every pair of events has a fixed causal order.  Quantum causality relaxes this constraint, it allows for scenarios ``event A is in the causal past of event B” and ``event B is in the causal past of event A” to be in superposition. Although counter-intuitive, this is plausible in the light of general relativity—where the causal order is dynamic rather than fixed \cite{hardy2007towards}. We already know quantum theory is incompatible with local-realistic theories that assign definite, pre-existing values to physical observables \cite{bell64},  we can extend this idea to causal order and imagine processes that have \emph{indefinite} causal order.  Implicit in such a process is the superposition of the order of events. It is this superposition of the order of events—rather than indefinite cause-effect relationships— that has been demonstrated in experiments so far.  

Indefinite causal order is of interest to the foundational quantum physics community because a fixed causal order might not be a feature of a theory that successfully combines general relativity and quantum physics. 
For a more pedagogical approach focusing on the foundational implications of quantum causality, curious readers are invited to read the review by \citet{Brukner_causality_review}.

From a pragmatic standpoint, indefinite causal order offers interesting extensions of conventional quantum Shannon theory \cite{Ebler_2018, chiribella2018indefinite, salek2018quantum} as well as computational\cite{chiribella09, araujo14, Ara_jo_2017} and communication complexity advantages \cite{Guerin2016}.  Our objective in this review is to provide a summary of this type of experiments without discounting other experiments that have broader foundational significance. Section II starts with a mathematical description of a process with indefinite causal order.  We give the example of a \emph{quantum switch}\cite{chiribella09}, a device that exhibits indefinite causal order. Section III focuses on how we can experimentally detect indefinite causal order with the use of a \emph{causal witness}\cite{araujo15}. Section IV discusses experiments that have shown communication advantages by exploiting indefinite causal order. A common requirement for all these experiments is to have two different carriers of information, one as a control of the order and one as a target to operate on. For Sections III and IV, we will broadly divide these implementations into two categories based on the degree of freedom that is used to control the order: those that use the path of a photon and those that use polarisation.  There is still an active discussion in the quantum causality community about the origin of the communication advantages \cite{Ebler_2018,salek2018quantum, chiribella2018indefinite} that have been claimed in recent experiments \cite{goswami2018communicating, Guo_2020}. We will summarise recent theoretical debates and comment on the relevance for  current experiments. Section V discusses computational advantages for three tasks: distinguishing commuting and anti-commuting operations, evaluation exchange game, and Hadamard promise problem.  Finally, Section VI gives a summary and outlook for the future. Note that in this review, we will often refer to ``causal order” simply as ``order” for brevity.
\begin{figure}[h]
\begin{center}
\includegraphics[width=\columnwidth]{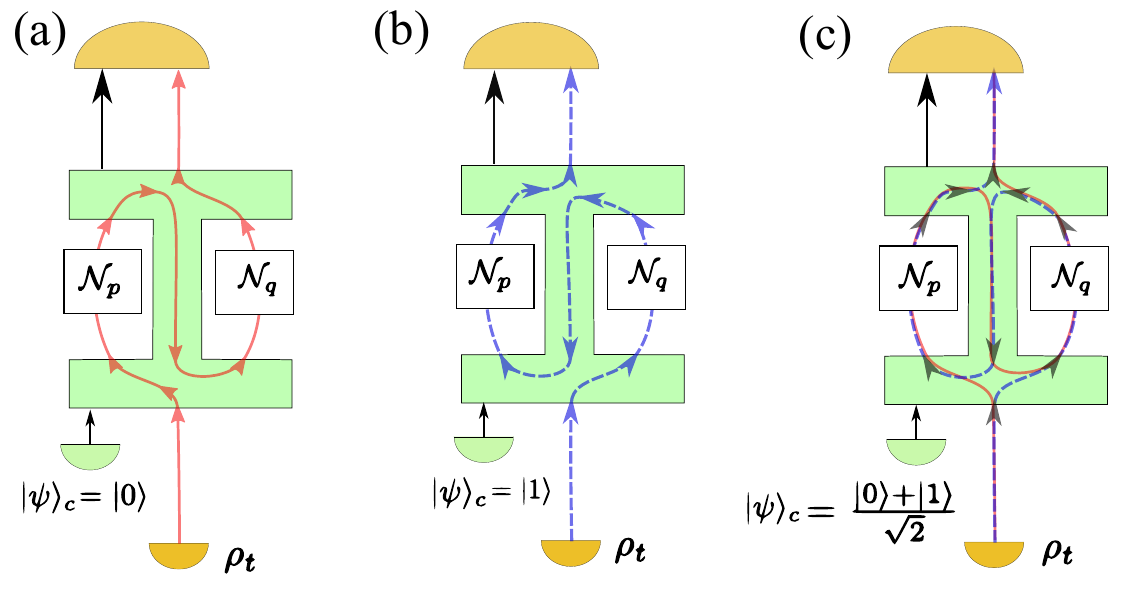}
\setlength{\belowcaptionskip}{-20pt}
\caption{Indefinite causal order.  The order that two individual operations, $\pazocal{N}_p$ and $\pazocal{N}_q$, act on a target qubit is controlled by a control qubit $\ket{\psi}_c$. (a) When the control qubit $\ket{\psi}_{c}{=}\ket{0}$, the operations have a definite order $\pazocal{N}_q {\circ} \pazocal{N}_p$. (b) When the control qubit $\ket{\psi}_{c}{=}\ket{1}$, the order is $\pazocal{N}_{p} {\circ} \pazocal{N}_{q}$.   (c) When the control is in a superposition, $\ket{\psi}_{c}{=}(\ket{0}{+}\ket{1})/\sqrt{2}$, the operations have an indefinite order. Reprinted with permission from K. Goswami \emph{et al}., ``increasing communication capacity via superposition of order," Phys. Rev. Res. \textbf{2}, 033292 (2020) (Ref.~\onlinecite{goswami2018communicating}). Copyright 2020 by APS.}
\label{fig:switch}
\end{center}
\end{figure}

\section{Indefinite  causal order} 
In conventional quantum mechanics, events take place in a definite order. We can relax this assumption and consider a scenario where the order of the events is not fixed or well-defined. The uncertainty in the order of events can be attributed to classical randomness \cite{Lamport78}---when the order is simply a mixture of the different possible orders---or quantum coherence---when the order of events is a superposition of the different possible orders \cite{oreshkov12, chiribella09}. A scenario for which the latter is true is said to have an \emph{indefinite causal order}.

To represent the events, one performs \emph{quantum operations} incorporating any combination of state preparation, transformation or measurement. The most general operation, performed by a party $X$, is given by a completely positive (CP) map $\pazocal{N}:X_I{\rightarrow}X_O$ --- a transformation from an input space $X_I {\equiv} \pazocal{L}\left(\pazocal{H}^{X_I}\right)$ to an output space $X_O {\equiv} \pazocal{L}\left(\pazocal{H}^{X_O}\right)$. Here, $\pazocal{L}\left(\pazocal{H}\right)$ denotes the space of bounded operators over a Hilbert space $\pazocal{H}$. It is convenient to represent a CP map $\pazocal{N}$ as a positive semidefinite matrix $N^{X_IX_O}{=}[\mathbb{1}{\otimes}\pazocal{N}(|\mathbb{1}{\rangle}\!{\rangle}\!{\langle} \!{\langle}{\mathbb{1}})]^{\pazocal{T}}{\in}X_I{\otimes}X_O$, as given by the \emph{Choi-Jamio{\l}kowski (CJ) isomorphism}~\cite{choi75,jamio72}. Here $\pazocal{T}$ is the transpose, and $\Ket{\mathbb{1}}=\sum_i\ket{i}\!\ket{i}$ is the unnormalised maximally entangled state where $\{\ket{i}\}$ is an orthonormal basis in $X_I$. For a deterministic quantum operation, e.g. a quantum channel, the CP map $\pazocal{N}$ is trace preserving (TP), and the corresponding CJ isomorphism satisfies $\tr_{X_O}N^{X_IX_O}{=}\mathbb{1}^{X_I}$, with $\mathbb{1}^{X_I}$ being the identity matrix in the input Hilbert space $X_I$.    

To encode the underlying causal structure between different quantum operations, the framework of \emph{process}, represented by the \emph{process matrix}, was developed \cite{oreshkov12}. Based on the process framework, the probability to realise the maps $\left\{\pazocal{N}_1, \pazocal{N}_2, \dots\right\}$ in an experiment, performed by the parties $\left\{A, B,\dots\right\}$, is given by the \emph{generalised Born rule}~\cite{gutoski06,chiribella09b,oreshkov12,shrapnel2017} 
\begin{equation}
P(\pazocal{N}_1,\pazocal{N}_2,\dots)=\tr\left[\left(N_1^{A_IA_O}\otimes N_2^{B_IB_O} \otimes \dots\right) W\right],
\label{Born}
\end{equation}
where $W {\in} A_I {\otimes} A_O {\otimes} B_I {\otimes} B_O {\otimes} \dots$ is the \emph{process matrix}. It is a positive semidefinite matrix which provides a full specification of the possible correlations that can be observed between the input and output systems. We call a process, $W_{\mathrm{sep}}$, \emph{causally separable} when it can be represented as a convex combination of different permutations of events---$W_{\mathrm{sep}}{=}\sum_{\upsilon}p_{\upsilon}W^{\upsilon}$, where $W^\upsilon$ is the corresponding process for the permutation $\upsilon$ and $p_{\upsilon}$ is the probability of occurrence of the permutation $\upsilon$. Every other process that does not decompose in this way is said to be \emph{causally nonseparable}, i.e. they exhibit indefinite causal order.

One can physically realise such a coherent-control protocol, for example in a \emph{quantum switch}\cite{chiribella09,chiribella12,colnaghi11}.  A quantum switch is comprised of two quantum systems---target, $\rho_t$ and control, $\rho_c{=}\ketbra{\psi}{\psi}_c$.  When the control is two-dimensional---a qubit---we consider two complete positive (CP) operations $\pazocal{N}_p$ and $\pazocal{N}_q$ acting on the target system. The control determines the order of the operations on the target: e.g. when $\ket{\psi}_c {=} \ket{0}$, the resulting operation is $\pazocal{N}_q {\circ} \pazocal{N}_p$, and for $\ket{\psi}_c {=} \ket{1}$ the resulting operation is $\pazocal{N}_p {\circ} \pazocal{N}_q$  (see Fig.~\ref{fig:switch}). The action of the quantum switch is to transform two individual operations $\pazocal{N}_p$ and $\pazocal{N}_q$ to a new operation $T(\pazocal{N}_p, \pazocal{N}_q){=}\ketbra{0}{0}_c  {\otimes} \pazocal{N}_q {\circ} \pazocal{N}_p{+}  \ketbra{1}{1}_c{\otimes} \pazocal{N}_p {\circ} \pazocal{N}_q$. When the operations $\pazocal{N}_p$ and $\pazocal{N}_q$ are quantum channels, we can give the equivalent Kraus representation \cite{Kraus} as
\begin{align}
   &T({\pazocal{N}_p},{ \pazocal{N}_q})_{ij} {=} \ketbra{0}{0}_c  {\otimes} {K^{(q)}_i}{K^{(p)}_j}{+}  \ketbra{1}{1}_c{\otimes} {K^{(p)}_j}{K^{(q)}_i}.
    \label{eq:switch}
\end{align}
Here, $\{K^{(p)}_i\}$ are the Kraus operators of $\pazocal{N}_p$---$\pazocal{N}_p(\rho_t){=\sum_{i}K^{(p)}_i\rho_t K^{(p){\dagger}}_i}$. Similarly, $\{K^{(q)}_i\}$ are the Kraus operators of $\pazocal{N}_q$. With this, the joint input state $\rho_c {\otimes} \rho_t$ transforms to the output state $T(\pazocal{N}_p, \pazocal{N}_q)(\rho_c \otimes \rho_t){=}\sum_{i,j}T({\pazocal{N}_p},{ \pazocal{N}_q})_{ij}(\rho_c {\otimes} \rho_t)T({\pazocal{N}_p},{ \pazocal{N}_q})_{ij}^{\dagger}$. Thus the coherence in the control system translates to coherence in the order of operations. For example, a control qubit in superposition, $\ket{\psi}_c{=}(\ket{0}_c{+}\ket{1}_c)/\sqrt{2}$ results in superposition of two different orders.

It is interesting to see the relationship between the switch operation $T(\pazocal{N}_p, \pazocal{N}_q)$ and the corresponding process formalism. In terms of the process formalism, a quantum switch consists of three parties---$A$ corresponding to the operation $\pazocal{N}_p$, $B$ corresponding to $\pazocal{N}_q$, and $C$ is the party at the output of the quantum switch who performs, in general, a joint operation occurring after both $A$ and $B$. There are two possible situations---$A$ being in the causal past of $B$ ($A\prec B\prec C$), and $B$ is in the causal past of $A$ ($B\prec A\prec C$). The process matrix of the quantum switch is $W={\ketbra{w}{w}}$, where 

\begin{align}
    &\ket{w}{=}\frac{1}{\sqrt{2}}\left(\ket{0}_c^{C_I^c}\ket{w^{A\prec B\prec C}}{+}\ket{1}_c^{C_I^c}\ket{w^{B\prec A\prec C}}\right), \label{Eq:switch_process} \\
    &\ket{w^{A\prec B\prec C}}{=}\Ket{\mathbb{1}}^{\mathcal{T}A_I}\Ket{\mathbb{1}}^{A_OB_I}\Ket{\mathbb{1}}^{B_OC_I^t}, \label{Eq:A_to_B}\\
    &\ket{w^{B\prec A\prec C}}{=}\Ket{\mathbb{1}}^{{\mathcal{T}}B_I}\Ket{\mathbb{1}}^{B_OA_I}\Ket{\mathbb{1}}^{A_OC_I^t}.\label{Eq:B_to_A}
\end{align}
Here, ${\mathcal{T}}$ is the Hilbert space associated with the target $\rho_t$, $C_I^c$ and $C_I^t$ denote party $C$'s input Hilbert space of the control and target systems respectively. Here, the dimension of all the systems in Eqs.~\eqref{Eq:A_to_B} and~\eqref{Eq:B_to_A} are the same, effectively the dimension of the target system. With this we can represent the output of the quantum switch,  $T(\pazocal{N}_p,\pazocal{N}_q)(\rho_c{\otimes}\rho_t)$ as
\begin{align}
 &T(\pazocal{N}_p,\pazocal{N}_q)(\rho_c{\otimes}\rho_t) \nonumber \\
 &{=}\tr_{\mathcal{T}A_IA_OB_IB_O}\bigg[\bigg\{\left(\rho_c^{C_I^c}{\otimes}\rho_t^{\mathcal{T}}\right)^\pazocal{T}{\otimes}N_p^{A_IA_O}{\otimes}N_q^{B_IB_O}\bigg\}.W \bigg].
\end{align}
Detecting whether a process has indefinite causal order is similar to detecting entanglement in nonseparable quantum states. We discuss this in the next section.

\begin{figure}[!t]
\begin{center}
\includegraphics[width=0.86\columnwidth]{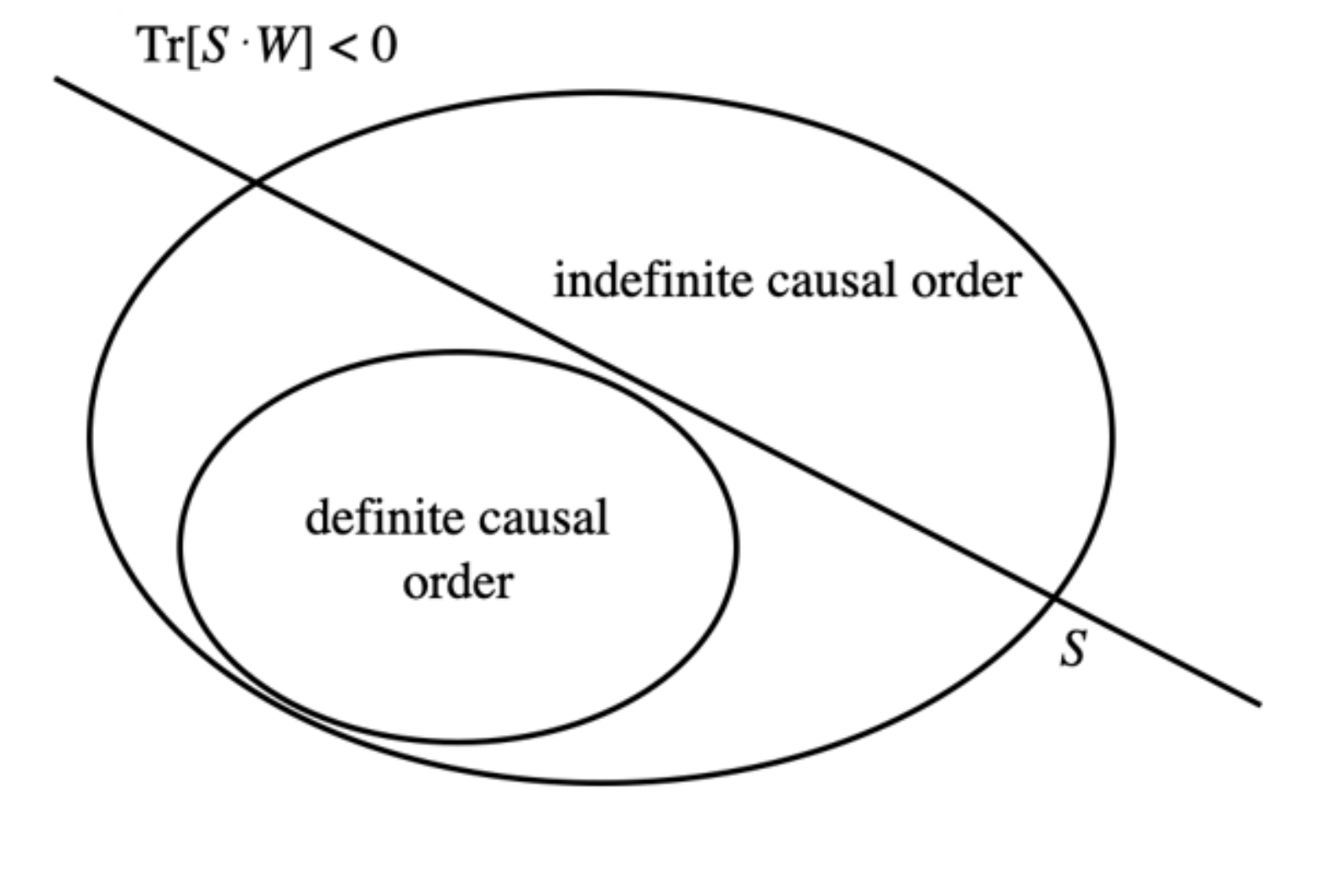}
\vspace{-2mm}
\end{center}
\caption{Causal witness $S$. A causal witness separates the processes that exhibit indefinite causal order from those that do not. For a process $W$, with indefinite causal order, there exists a causal witness, namely an operator $S$ which results Tr$[S\cdot W]<0$. For any non-separable process $W_{\mathrm{sep}}$ results Tr$[S\cdot W_{\mathrm{sep}}]{\geq}0$.}
\label{fig:causal_witness_experiment}

\end{figure}

\section{Causal witness}
In the same way that some nonseparable states can violate a Bell inequality \cite{bell64} in a device-independent manner, some causally nonseparable processes can violate a \emph{causal inequality} \cite{oreshkov12}. Violation of a causal inequality is impossible with space-like and time-like correlations \cite{oreshkov12}, and there is currently no known implementation of a violation that is experimentally feasible in our laboratories.   

The quantum switch---although causally nonseparable---does not violate a causal inequality. In the same way that an entanglement witness \cite{guhne09} can be used to detect entanglement in nonseparable quantum states that do not violate a Bell inequality, a device-dependent quantity called \emph{causal witness} can be used to confirm whether a process is causally  nonseparable \cite{araujo15}.
The formulation of the causal witness is governed by the separating hyperplane theorem \cite{rockafellar70}, which states that it is always possible to find a hyperplane that separates two disjoint, closed-convex sets (see Fig.~\ref{fig:causal_witness_experiment}). Owing to the fact that causally-separable process matrices form a closed-convex set, for a particular causally non-separable process $W$ it is possible to construct a witness $S$ such that
\begin{align}
    \tr[S\cdot W] < 0. 
\end{align}
\noindent The same operator $S$, for causally separable processes $W_\mathrm{sep}$, results $\tr [S\cdot W_{\mathrm{sep}}] \geq 0$. 

The value of $\tr [S\cdot W]$ can be obtained experimentally after decomposing $S$ into different operators representing different CP maps (which define the events $A,B,C$), and using Eq.~\eqref{Born} to write $\tr [S\cdot W]$ as a combination of the joint probabilities of these maps to be realised.  There have been two experiments demonstrating that the causal witness can indeed detect indefinite causal order, differing in their implementations of the quantum switch \cite{rubino2017,Goswami_2018}.

A quantum switch requires two systems, one to act as control and one as target. Although the description of a quantum switch is agnostic to the physical platform,  experiments so far have been photonic in nature. One degree of freedom (DOF) is used as a control and another DOF in the same particle is used as a target. Mathematically, this is no different from the case of the control and target systems residing in separate particles. 
In order to describe the experiments measuring the causal witness, we classify the quantum switch implementations based on the DOF used for control: either path of the photon \cite{rubino2017}, or polarisation\cite{Goswami_2018}. In the former\cite{rubino2017}, event $A$ is represented by a set of \emph{measure-prepare operations} on the polarisation ($\pazocal{N}_p$)---four different measurement operations on the incoming polarisation and re-preparation of the polarisation in one of the three pre-assigned states. The events associated with $B$ are ten different unitary operations on the polarisation ($\pazocal{N}_q$). In the second experiment \cite{Goswami_2018}, both of the events $A$ (operations $\pazocal{N}_p$) and $B$ (operations $\pazocal{N}_q$) are chosen from six unitary operations acting on the transverse spatial mode of the photon. In both cases, the event $C$ corresponds to a measurement operation on the control system ($\pazocal{C}$). With this, one can obtain the value of $\tr [S\cdot W]$ experimentally after decomposing $S$ into CP maps associated with the events $A,B,C$,

\begin{align}
    S {=}\sum_{a,b,c,d} \alpha_{a,b,c,d} \rho_t^{(a)}{\otimes}{\pazocal{N}_p}^{(b)}{\otimes} {\pazocal{N}_q}^{(c)}{\otimes}\pazocal{C}^{d}.
    \label{map}
\end{align}

The superscipts $a,b,c,$ denote different choices for the operations on the target system and $d$ labels the final measurement operation on the control. Using the decomposition in Eq.~\ref{map} and Eq.~\ref{Born},  $\tr [S\cdot W]$ is evaluated as a combination of the joint probabilities of the maps. If the causal witness $\langle S \rangle := \tr[S\cdot W]$ is negative we can conclude the quantum switch indeed exhibits indefinite causal order, i.e. the order of events $A$ and $B$ is in a superposition.

\subsection{Experiment with path as control}

\begin{figure}[!t]
\begin{center}
\vspace{-5mm}
\includegraphics[width=0.8\columnwidth]{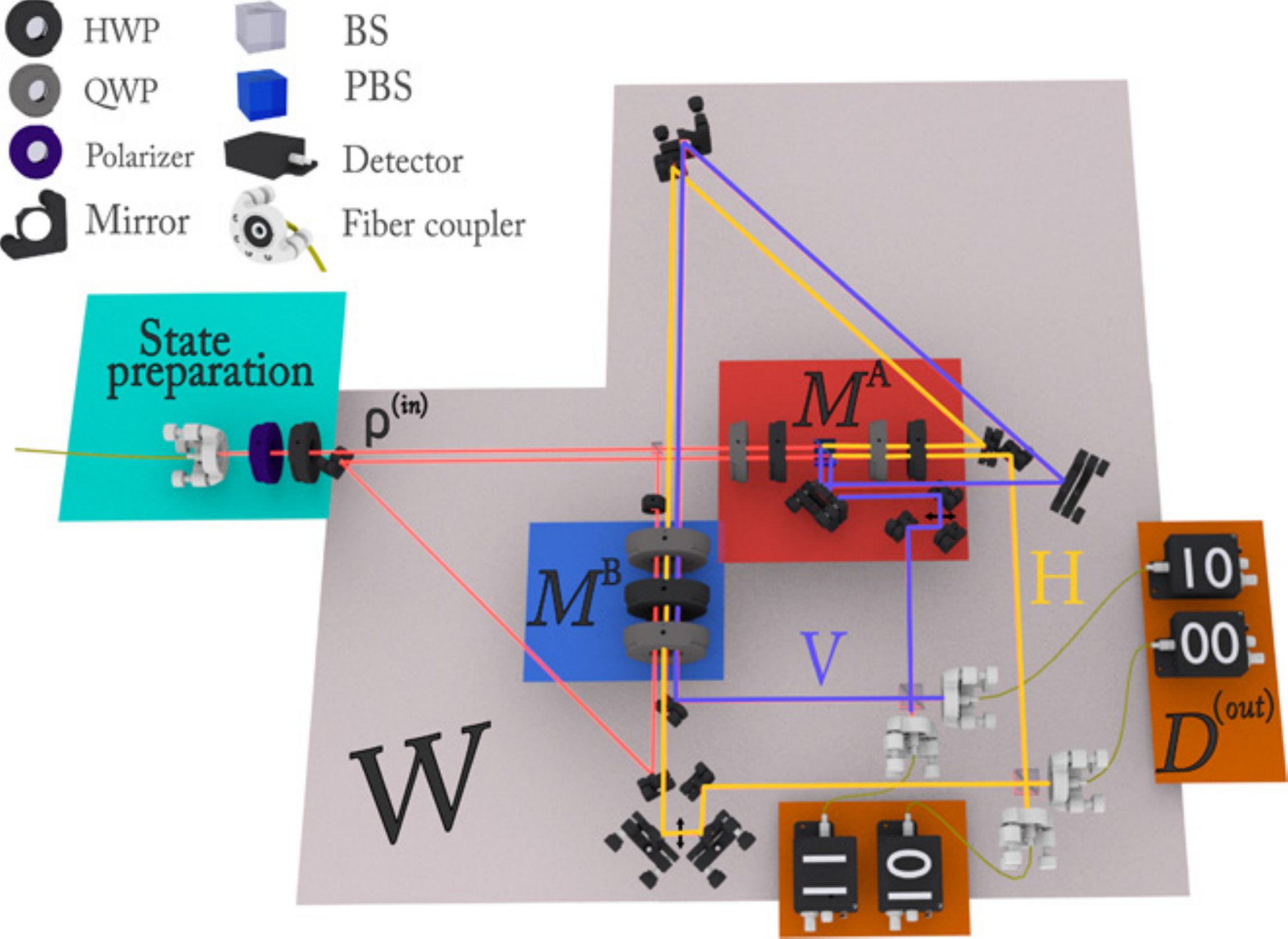}
\vspace{-2mm}
\caption{Quantum switch with path as control. A pair of single photons is produced via SPDC (not shown). One photon serves as a herald while the other is sent to the quantum switch. The target system is encoded in polarisation and is prepared using waveplates. After state preparation, the photon is sent to a 50/50 beamsplitter which prepares the control in a superposition of two paths: one where $M^A$ comes first and one where $M^B$ comes first. Note that there are two interferometers---the photon impinges on a different spot on the waveplate depending on the order of the operations.  The yellow (purple) path corresponds to the case when the photon entering $M^A$ is horizontally-(vertically-) polarised, denoted by the first digit in the detectors being 0(1). The second digit in the detectors correspond to the final measurement outcome where 0 (1) indicates that a photon exited from a horizontally-drawn (vertically-drawn) port.   Reprinted with permission from G. Rubino \emph{et al.}, ``Experimental verification of an indefinite causal order," Sci. Adv. \textbf{3}, e1602589 (2017) (Ref.~\onlinecite{rubino2017}). Copyright 2017 by AAAS.
}
\label{fig:vienna_witness_experiment}
\end{center}
\end{figure}

One way to implement a quantum switch is to use the path of photons as control for the order of operations. In Fig.\ref{fig:vienna_witness_experiment}, the authors used type II spontaneous parametric down-conversion (SPDC) to generate 790 nm single photons \cite{rubino2017}. One photon was used as a herald and the other photon was used as input to the interferometer shown in Fig.\ref{fig:vienna_witness_experiment}. The target system (polarisation) is prepared by a set of waveplates, and then sent to a 50/50 beamsplitter (BS).  The beamsplitter sets the state of the control system (path) to be in the superposition of two paths---one first going to $M^A$ (corresponding to operations $\pazocal{N}_p$ in Eq. \ref{map}) and one first going to $M^B$ (corresponding to operations $\pazocal{N}_q$ in Eq. \ref{map}).  The operations $M^A$ in Fig. \ref{fig:vienna_witness_experiment} are \emph{measure-prepare} operations on polarisation, consisting of four possible measurements---implemented by a  set of waveplates followed by a polarising beamsplitter---and three possible state re-preparations---implemented by a set of waveplates. The operations $M^B$ in Fig. \ref{fig:vienna_witness_experiment}  are chosen from a set of ten unitary operations on polarisation, realised by a train of quarter waveplate (QWP), half waveplate (HWP) and a second QWP. After the two operations, the photon was detected in any of four output ports labelled as in Fig. \ref{fig:vienna_witness_experiment} where the first bit denotes the result of the measurement in $M^A$ and the second bit indicates the output port of the control qubit. The experiment was repeated with three different input target states each of multiple copies to build statistics.  The theoreticaly predicted value of the witness is $-0.2842$, the experimentally obtained value is $-0.202{\pm}0.029$.  Because of the separate paths for each of the orders, the dominant source of error in this experiment is the dephasing of the control qubit due to path length difference and other systematic experimental imperfections. The two paths should remain interferometrically stable: as the imbalance in the path length grows, the value of the causal witness increases until it becomes greater than zero for visibilities below $\sim66\%$.

\subsection{Experiment with polarisation as control}
\begin{figure}[!t]
\begin{center}
\includegraphics[width=0.8\columnwidth]{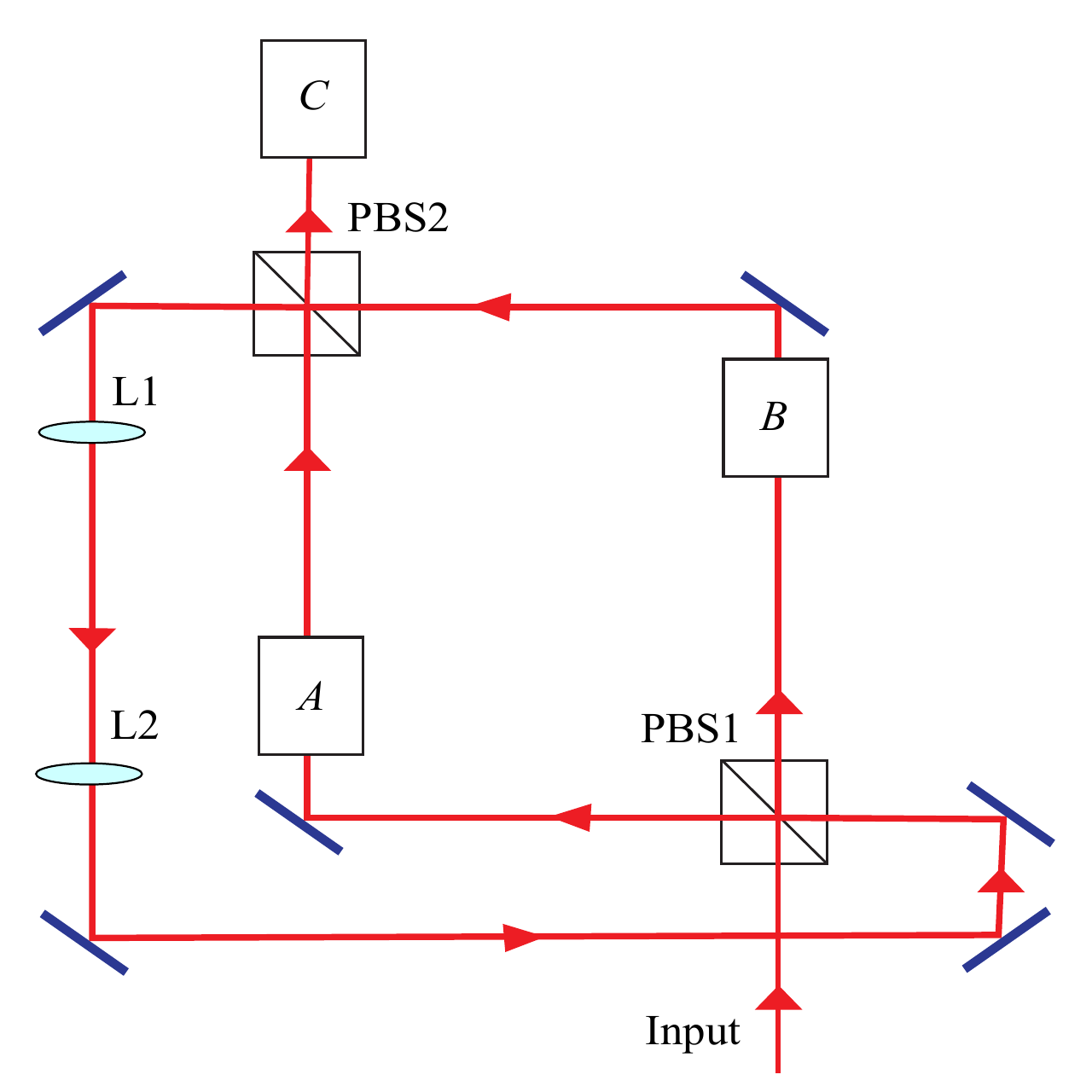}
\vspace{-2mm}
\setlength{\belowcaptionskip}{-15pt}
\caption{Quantum switch using polarisation as control.  The polarising beamsplitter PBS1 routes the photon into either events A or B, which realise unitary operations $\pazocal{N}_p$ and $\pazocal{N}_q$ respectively. Event C is a polarisation measurement which determines the Stokes parameter of the photon in the diagonal/anti-diagonal basis. Lenses L1 and L2 are used as a telescope to ensure mode-matching. Reprinted with permission from K.   Goswami \emph{et al.}, ``Indefinite causal order in a quantum switch," Phys. Rev. Lett. \textbf{121}, (2018)(Ref.~\onlinecite{Goswami_2018}). Copyright 2018 by APS.
 }
\label{fig:indefinite causal order UQ setup}
\end{center}
\end{figure}

\begin{figure}[!t]
\begin{center}
\includegraphics[width=\columnwidth]{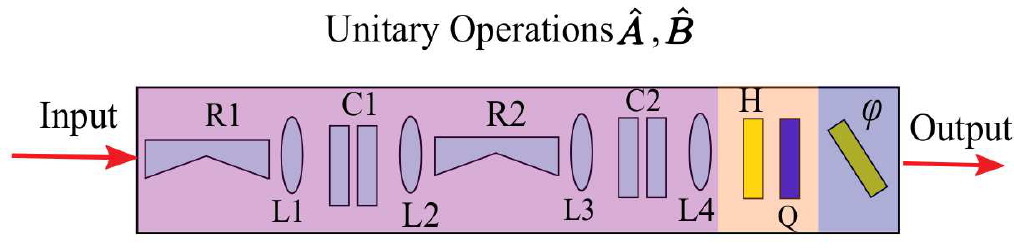}
\vspace{-2mm}
\caption{Realising unitary operations in the transverse spatial mode. The set of six Pauli operations $\{\mathbb{1}, \sigma_x, \sigma_y, \sigma_z, P{=}(\sigma_y{+}\sigma_z)/\sqrt{2}, Q{=}(\sigma_x{+}\sigma_z)/\sqrt{2}\}$ were implemented using  inverting prisms (R), cylindrical lenses (C), and spherical lenses (L). The prisms rotate the incoming transverse mode, the cylindrical lenses give a $\pi/2$ relative phase shift, and the spherical lenses or mode-matching. The half-waveplates (H), quarter-waveplates (Q), and phase-plate ($\varphi$) are used to correct polarisation changes caused by reflections in the prisms and $\phi$ represents a phase plate. Reprinted with permission from K.   Goswami \emph{et al.}, ``Indefinite causal order in a quantum switch," Phys. Rev. Lett. \textbf{121}, (2018) (Ref.~\onlinecite{Goswami_2018}). Copyright 2018 by APS.}
\label{fig:indefinite causal order setup unitary}
\end{center}
\end{figure}

In another implementation of the quantum switch\cite{Goswami_2018}, photon polarisation was used as the control qubit for the order of operations that act on the shape--- transverse spatial mode---of the photon.   Fig.~\ref{fig:indefinite causal order UQ setup} shows the experiment schematic, with $A$ and $B$ corresponding to the operations $\pazocal{N}_p$ and $\pazocal{N}_q$ in Eq. \ref{eq:switch}. A  narrowband (with a coherence length much longer than the length of the interferometer) 795 nm weak coherent laser serves as the input to the first polarising beamsplitter (PBS1) in Fig.~\ref{fig:indefinite causal order UQ setup}. A vertically(horizontally) polarised photon undergoes operations $A$($B$) first followed by  operations $B$($A$). When the control qubit is set to diagonal polarisation, the order of the operations are in a superposition. The target qubit in the transverse spatial mode was set to a first-order Hermite-Gaussian mode.  For the causal witness, the operations in $A$ and $B$ were chosen from a set of six Pauli unitaries $\{\mathbb{1}, \sigma_x, \sigma_y, \sigma_z, P{=}(\sigma_y{+}\sigma_z)/\sqrt{2}, Q{=}(\sigma_x{+}\sigma_z)/\sqrt{2}\}$. They were implemented using a train of rotating prisms and cylindrical lenses, as shown in Fig.~\ref{fig:indefinite causal order setup unitary}. For each of the two operations, the polarisation was measured in the diagonal/anti-diagonal basis.  The results were combined as in Eq. \ref{map} and a causal witness value of $-0.171{\pm}0.009$ was obtained.  This value is different from the ideal value of the causal witness which is $-0.248$. Unlike in Fig. \ref{fig:vienna_witness_experiment}, the two possible orders of $A$ and $B$ in Fig.~\ref{fig:indefinite causal order UQ setup} share a common path so path length difference is not an issue in this kind of quantum switch implementation. However, the shape of light is highly sensitive to any imperfections in the transverse plane, e.g. imperfect surface flatness of the optical elements and non-ideal mode matching.  The degradation of the visibility that results from these imperfections lead to an increase in the value of the causal witness. Despite the deviations from the ideal case, the causal witness is still negative indicating indefinite causal order in the quantum switch.  

\section{Bell's theorem for temporal order}
The framework of causal witness presented in the previous section is theory-dependent in that it assumes that quantum mechanics is true. However, a large class of \emph{generalised probabilistic theories} which respects local-realism---and therefore satisfies Bell inequality---could potentially explain the experimental outcomes without assuming an indefinite causal structure. A Bell inequality to distinguish between these generalised probabilistic theories and those that exhibit indefinite causal order was introduced by Zych et al. \cite{zych2017bell} 
 An experiment that violates such a Bell inequality was implemented by Rubino et al.~\cite{rubino2017b} It consists of two \emph{entangled quantum switches}  with separable target systems in polarisation and control systems entangled in path. 

\begin{figure*}[!t]
\begin{center}
\includegraphics[width=1.7\columnwidth]{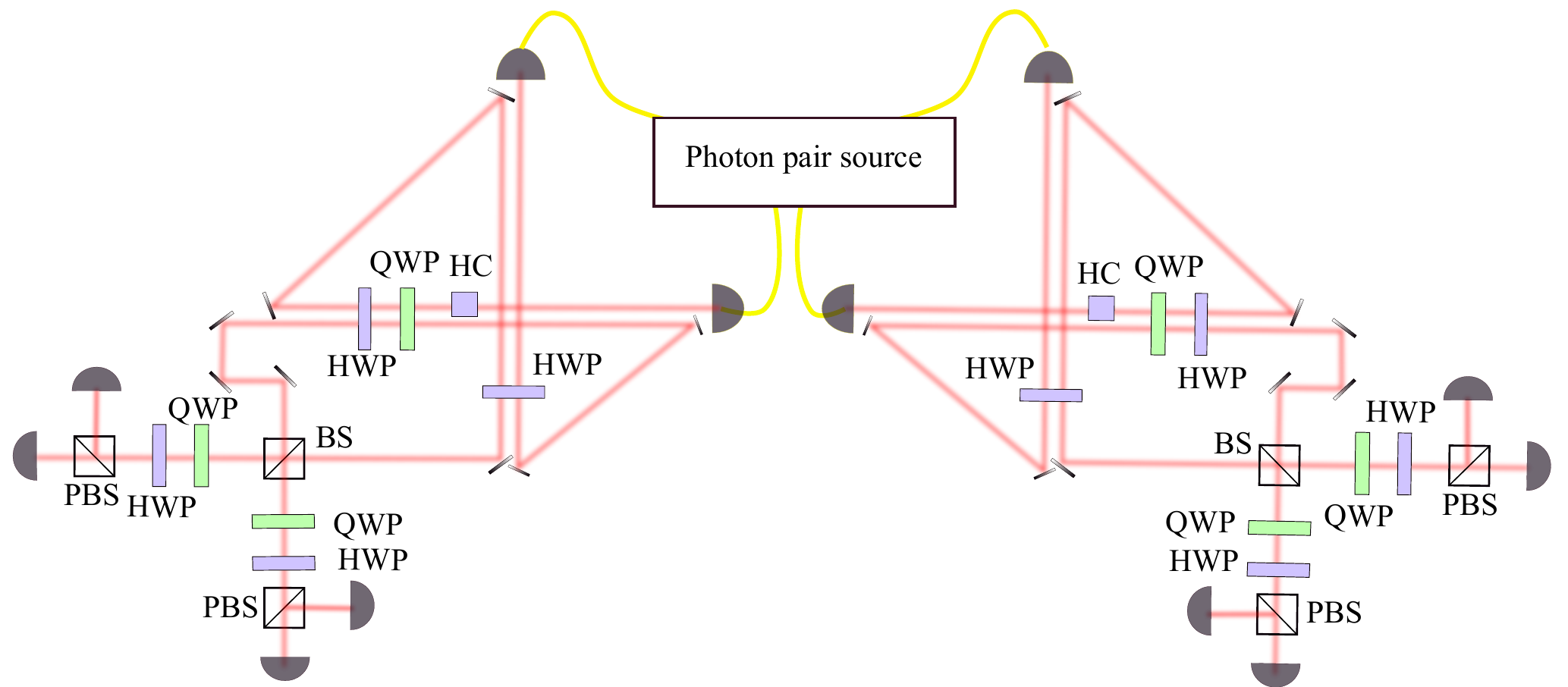}

\caption{Experimental violation of Bell's theorem for temporal order, as described in Ref.~\onlinecite{rubino2017b}. (a)  Two quantum switches with polarisation as target and path has control were used to show a violation. (PBS: polarising beamsplitter, QWP:quarter- waveplate, HWP:half-waveplateHC, HC:half-waveplate compensation, BS: beamsplitter. All output ports of the beamsplitters are monitored by single-photon detectors.)  }
\label{fig:bell setup}
\end{center}
\end{figure*}

\citet{rubino2017b} use the setup in Fig.~\ref{fig:bell setup} to violate the following no-go theorem:
No states, set of transformation, or measurements violate Bell's inequality given 
(1) the initial joint state of the target system does not violate Bell's inequality; (2) the laboratory operations are local transformation, i.e. they do not increase the amount of violation of Bell's inequality; (3) the order of operations on the two target systems are pre-defined.

When assumptions 1 and 2 are both satisfied, the violation of Bell's inequality can only be associated with indefinite causal order. 

 To ensure the validity of assumption (1) a state tomography of the input joint target systems was performed. They obtained the polarisation state $\ket{HH}(\ket{00})$ with a fidelity of $0.935{\pm}0.004$ with  and a concurrence of $0.001{\pm}0.010$. Morevoer, a probability measurement on the joint target systems showed that the joint probability distribution was the product of the marginals, suggesting that the initial joint target systems are reasonably separable.

 Assumption (2) is enforced by keeping the two operations  spatially separated in the optical setup, thus eliminating the possibility of entangling operations. Moreover, the local operations cannot increase the amount of violation of Bell's inequality. This was supported by disentangling both initial  target-control systems and performing corresponding probability measurement at the output. The probability distribution turned out to be the product of the marginals.  
 
With the first two assumptions verified, Rubino et al.~\cite{rubino2017b} violated the Bell inequality using two quantum switches, $T_1$ and $T_2$, in a two-loop Mach-Zehnder interferometer as shown in Fig.~\ref{fig:bell setup} . They start with separable target systems , $\ket{0}_t^{T_1}\ket{0}_t^{T_2}$, and control systems entangled in path,  $(\ket{0}_c^{T_1}\ket{0}_c^{T_2}-\ket{1}_c^{T_1}\ket{1}_c^{T_2})/\sqrt{2}$.  The entanglement of the initial joint control system was verified with a Bell test resulting in a Bell parameter of $2.58{\pm}0.09$. The local operations applied were $U_A^{T_1}{=}U_A^{T_2}{=}\sigma_z$ and $U_A^{T_1}{=}U_A^{T_2}{=}(\mathbb{1}{+}i\sigma_x)/\sqrt{2}$, implemented via a set of waveplates. At the output, the authors measured the control systems in the diagonal/anti-diagonal basis via a 50-50 beamsplitter. This made the resulting target systems maximally entangled, $(\ket{l}_t^{T_1}\ket{l}_t^{T_2}-\ket{r}_t^{T_1}\ket{r}_t^{T_2})/\sqrt{2}$ with $\ket{l}{=}(\ket{0}{+}i\ket{1})/\sqrt{2}$ and $\ket{r}{=}(\ket{0}{-}i\ket{1})/\sqrt{2}$. A state tomography of the output joint target system gave a fidelity  of $0.922{\pm}0.005$ and a concurrence of $0.95{\pm}0.01$. The Bell test on the joint output target systems resulted in a Bell parameter of $2.55{\pm}0.08$. This violation of the Bell inequality means that assumption (3) is invalid, thus demonstrating indefinite causal order in a theory-independent manner.

\section{Application in quantum Shannon theory}
Aside from foundational implications, indefinite causal order brings about some practical advantages. We first focus on an advantage for communication: communicating through noisy channels. Optimising information transmission through noisy channels is the primary goal of information theory. Quantum physics played a major role in this endeavour by opening up new possibilities---such as unprecedented security---when the information carriers are themselves quantum\cite{Wilde13, Yin2020}. A quantum information theory that allows for indefinite causal order opens up further possibilities \cite{Ebler_2018,salek2018quantum, chiribella2018indefinite}, one example is being able to communicate through completely noisy channels. Several works\cite{Ebler_2018,salek2018quantum, chiribella2018indefinite} have shown that when two noisy channels, 
say $\pazocal{N}_p$ and $\pazocal{N}_q$, are connected in an indefinite order, there is a significant improvement in both classical and quantum information capacity. We first discuss the theoretical aspects of this communication advantage and then proceed to discuss two experiments that demonstrate these ideas.

 A standard way of modelling a noisy quantum channel is to exploit generalised Pauli channels. A Pauli channel is a probabilistic mixture of Pauli operations associated with three common errors: $\sigma_{1} {\equiv} \sigma_x$ (bit flip), $\sigma_{3} {\equiv} \sigma_z$ (phase flip), and $\sigma_{2} {\equiv} \sigma_y$ (combination of bit flip and phase flip). We denote as $\sigma_{0} {\equiv} I$, i.e. identity operation when there is no error. The action of such a channel $\pazocal{N}_p$ on a quantum state $\rho$ would be, 
 \begin{align}
\pazocal{N}_p(\rho){=}\sum_{i=0}^3p_{i}\sigma_{i}\rho_t\sigma_{i}^\dagger
 \label{channel}
 \end{align}

\noindent where $\sum_i p_{i} {=} 1$. Let us consider two such channels $\pazocal{N}_p$ and $\pazocal{N}_q$ acting on the target qubit $\rho_t$. Note that we keep the same notation---$\pazocal{N}_p$ and $\pazocal{N}_q$---as in the previous section,  only that we focus on the case that these are noisy channels rather than an arbitrary operation on the target system. Let us also consider a control qubit initially set to the state $\rho_{c}{=}|\psi\rangle\langle\psi|_c$, where $\ket{\psi}{=}\sqrt{\gamma}|0 \rangle {+}\sqrt{1{-}\gamma}|1 \rangle $. The total output state of the switch becomes\cite{goswami2018communicating} 

\begin{equation}
T[\pazocal{N}_{p},\pazocal{N}_{q}](\rho_c{\otimes}\rho_t){=}\left(\begin{array}{cc}
A&B\\
B&\tilde{A}\\
\end{array} \right),
\label{output_AB_Matrix}
\end{equation}
with, 
\begin{equation}
\begin{array}{l}
A{=}\gamma \pazocal{N}_q\circ \pazocal{N}_p(\rho_t),\\
\tilde{A}{=}(1-\gamma)\pazocal{N}_p\circ \pazocal{N}_q(\rho_t),\\
B{=}\sqrt{\gamma(1-\gamma)}(\epsilon_+(\rho_t)-\epsilon_-(\rho_t)),\\
  \end{array}
\label{A_and_B}
\end{equation}
where ($\epsilon_-$) $\epsilon_+$ represents an auxiliary trace non-preserving map  
\begin{align}
 \epsilon_+(\rho_t) &{=}\sum_{i=0}^3p_iq_i\, \rho_t + \sum_{i=0}^3 r_{0i}\,\sigma_i\rho_t\sigma_i^\dagger \label{commutator_channel}\\
 \epsilon_-(\rho_t) & {=}r_{12}\,{\sigma_3}\rho_t\sigma_3^\dagger {+}r_{23}\,\sigma_1\rho_t\sigma_1^\dagger {+} r_{31}\,\sigma_2\rho_t\sigma_2^\dagger ,
\label{anticommutator_channel}
\end{align}
with $r_{ij} {=} p_iq_j{+}p_jq_i$. Note that $\pazocal{N}_p{\circ} \pazocal{N}_q(\rho_t) {=} \epsilon_+(\rho_t)+\epsilon_-(\rho_t)  {=} \pazocal{N}_q{\circ }\pazocal{N}_p(\rho_t)$, which means any definite order of $\pazocal{N}_p$ and $\pazocal{N}_q$ will have the same effect on the target qubit. The presence of the nontrivial off-diagonal block matrix $B$ in the joint output state (Eq. \ref{output_AB_Matrix}) arises from the indefinite causal order and can lead to a communication advantage; $B$ vanishes in the case of definite order.  

Before proceeding, it is important to introduce the standard measure of classical and quantum information transmission in a channel, quantified via the classical capacity $C$ and the quantum capacity $Q$ respectively. The classical capacity is defined by the maximum rate of classical bits that can be transmitted over asymptotic uses of the channel (i.e. when the number of times a channel is used approaches infinity). A convenient measure to quantify the classical capacity is given by the Holevo capacity $\chi$, which for a channel $\pazocal{N}$ is defined by:
\begin{align}
    \chi(\pazocal{N}){=} \underset{p(m), \rho_m}{{\mathrm{max}}}  E\left(\sum_m p(m) \pazocal{N}( \rho_m)\right) {-} \sum_m p(m)E\left(\pazocal{N}(\rho_m)\right).
    \label{accessible_information}
\end{align}
Here, $E(.)$ is the von Neumann entropy, and the maximum is over the input ensembles $\{p(m), \rho_m\}$. Similarly, the quantum capacity $Q$ is the maximum rate that qubits can be transmitted over asymptotic uses of the channel. The one shot version of the quantum capacity is given by coherent information $Q_1$:
\begin{align}
    Q_1= \underset{\rho}{{\mathrm{max}}} \,  E\left(\pazocal{N}(\rho)\right){-}E\left[(
    \mathbb{1}{\otimes}\pazocal{N})\left(\ketbra{\psi_{\rho}}{\psi _{\rho}}\right)\right].
    \label{coherent_information}
\end{align}
Here, $\ket{\psi_{\rho}}$ is the purification of the input state $\rho$. The purification can be obtained by adding an ancillary system associated with an orthonormal basis $\{\ket{m}\}$ to the system of $\rho$. Considering $\rho$ in a diagonalised form $\sum_m\lambda_m\ketbra{\lambda_m}{\lambda_m}$, the purification becomes $\ket{\psi_\rho}{=}\sum_i\sqrt{\lambda_m}\ket{\lambda_m}\ket{m}$ \cite{chuang00}, where $\{\lambda_m\}$ and $\{\ket{\lambda_m}\}$ are the eigenvalues and the eigenvectors of $\rho$ respectively. The optimisation of the coherent information is done over all possible input states $\rho$\cite{Wilde13}.

\subsection{Experiment with polarisation as the control}
The schematic in Fig.~\ref{fig:indefinite causal order UQ setup} can be used to verify the communication advantage proposed by Ebler et al.\cite{Ebler_2018}.  The polarisation was used as the control for the order of Pauli operators. The Pauli operators act on the transverse spatial mode, which was initially set to a first-order Hermite-Gaussian mode. The Pauli operators were implemented using pairs of rotating prisms which were rotated depending on the operations\cite{goswami2018communicating}. 
All possible pairs the Pauli operations were implemented to be able to reconstruct the channels $\pazocal{N}_p$ and $\pazocal{N}_q$ with the relevant probability distributions \cite{goswami2018communicating}.

\begin{figure}[!b]
\begin{center}
\includegraphics[width=\columnwidth]{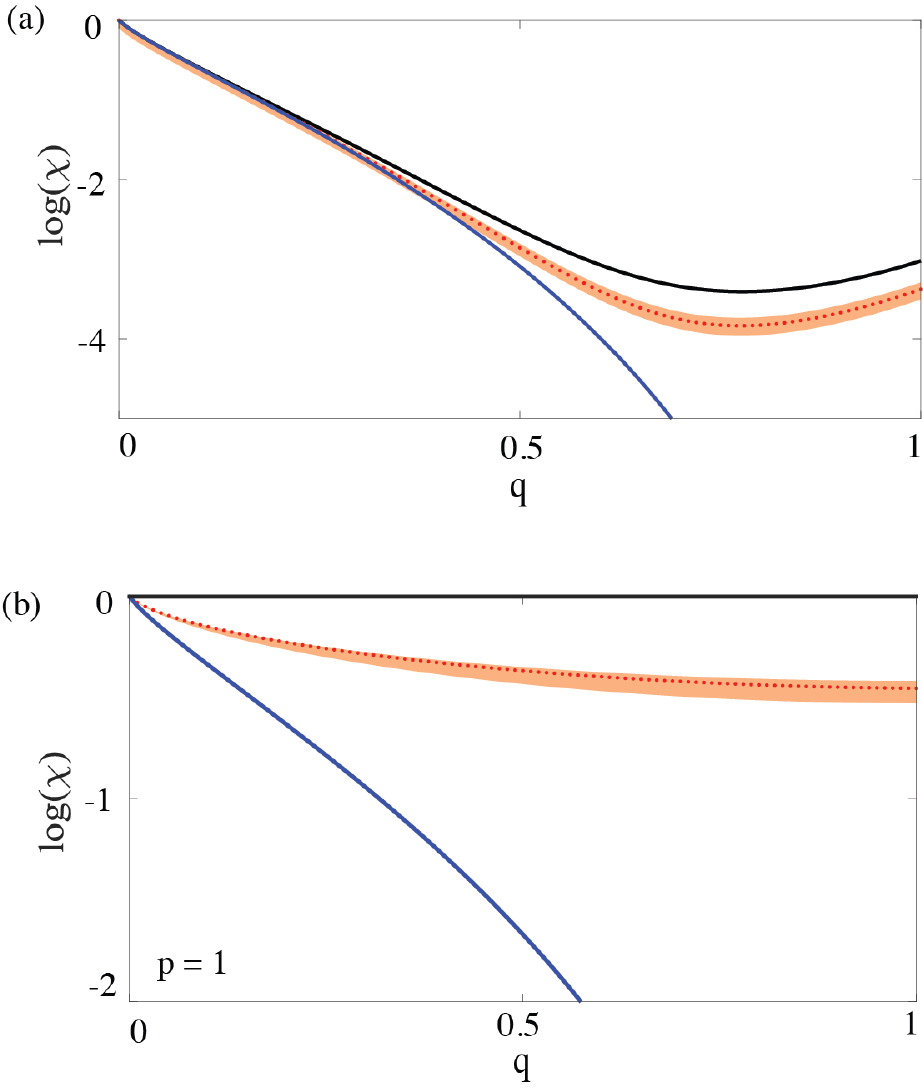}
\vspace{-2mm}
\caption{Logarithm of Holevo capacity $\chi$ when (a) both channels are depolarising channels and (b) one channel is $\sigma_z$ channel and the other is a depolarising channel, for varying depolarising strength $q$ (x-axis). The solid blue line is the capacity for definite causal order. The solid black line is the ideal capacity for the case of indefinite causal order. The red dots are experimentally-derived capacities with the orange shade showing the uncertainty due to non-ideal visibility. Reprinted with permission from  K. Goswami \emph{et al}., ``increasing communication capacity via superposition of order," Phys. Rev. Res. \textbf{2}, 033292 (2020) (Ref.~\onlinecite{goswami2018communicating}). Copyright 2020 by APS. }
\label{fig:depolarisation}
\end{center}
\end{figure}

\emph{Classical communication with two depolarising channels}
Ebler et al.\cite{Ebler_2018} proposed that two completely depolarising channels $\pazocal{N}(\rho){=}(\sum_i\sigma_i \rho \sigma_i$)/4 connected in an indefinite causal order is able to transmit a finite amount of information. This is particulaly surprising given that any definite ordered combination of these channels will scramble all the classical information. Fig. \ref{fig:depolarisation} shows the logarithm of the Holevo capacity as a function of depolarising strength $q$. The theoretical prediction for the Holevo capacity of two fully-depolarising ($q{=}1$) channels arranged in a superposition of order is $0.049$ bits.  As shown in Fig.~\ref{fig:depolarisation}.a, the experiment showed a transmission of $(3.4{\pm}0.2){\times}10^{-2}$ bits\cite{goswami2018communicating}.

\emph{Classical communication with one unitary and one depolarising channel.} A more dramatic example of information-theoretic advantage is possible when one channel is a Pauli operation and the other is a completely depolarising channel. A definite-ordered combination of these two channels result to a zero-capacity channel.  However, when these channels are arranged in an indefinite order, it is possible to achieve deterministic classical information transmission (solid black line in Fig. \ref{fig:depolarisation}.b) through the system even in the case of full depolarisation strength ($q{=}1$)\cite{goswami2018communicating}. As shown in Fig.~\ref{fig:depolarisation}.b, for all values of $q$ the experimental Holevo capacity of the indefinite-ordered case (orange curve) is significantly higher than the definite-ordered case (blue curve); for $q{=}1$, Holevo capacity is $0.64 {\pm}0.02$ bits.

\begin{figure*}[!t]
\begin{center}
\includegraphics[width=1.7\columnwidth, ]{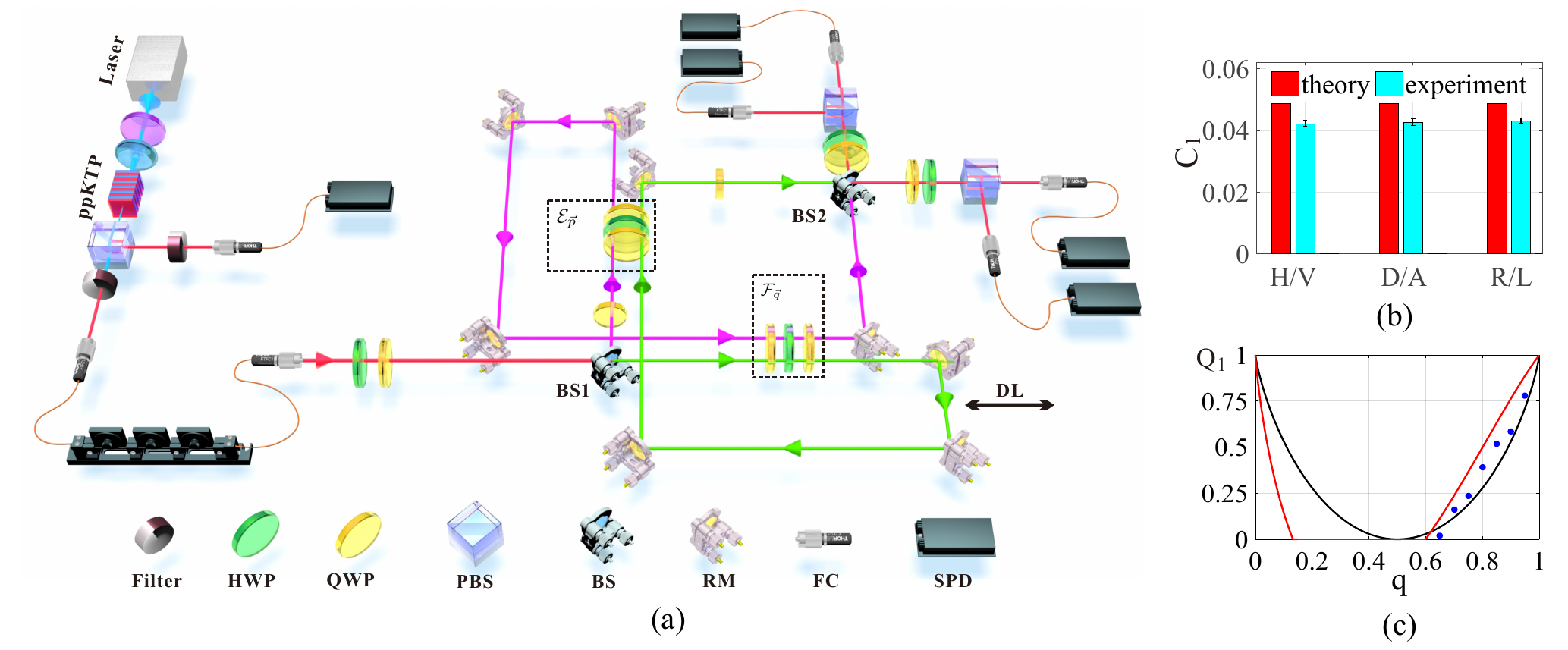}
\caption{(a) Experimental  setup \cite{Guo_2020}.   Pairs of photons were produced by spontaneous parametric down-conversion. One of the photons acted as a herald while the other is input to the interfereometer. The polarisation of the photon was used as target and the photon path acted as the control qubit. The noisy channels (here denoted by $\epsilon_{\Vec{p}}$ and $\pazocal{F}_{\Vec{q}}$) each consist of waveplates.  (HWP: half wave plate; QWP: quarter wave plate; PBS: polarizing beam splitter; BS: 50/50 beamsplitter; RM: mirror; FC: fiber coupler; SPD: single photon detector; DL: trombone-arm delay line.)(b) Experimental (cyan) classical capacity (denoted here by $C_1$) for the case of two fully-depolarising channels combined in indefinite order compared with ideal theoretical classical capacity (red). (c) One-shot coherent information $Q_1$ for the channels($\pazocal{N}_p$ and $\pazocal{N}_q$) as a function of q [0,1]. Blue dots show the experimental results, the corresponding theoretical prediction is shown  by  the  red  line.   For  comparison,  the  one-shot coherent information of the two dephasing channels combined in a definite order is also shown by the black curve. Reprinted with permission from Y. Guo \emph{et al.}, ``Experimental  transmission  of quantum information using a superposition of causal orders," Phys. Rev.Lett. \textbf{124}, (2020) (Ref. \onlinecite{Guo_2020}). Copyright 2020 by APS. }
\label{fig:guo setup}
\end{center}
\end{figure*}

\subsection{Experiment with path as the control}

To show communication advantage, Guo et al. implemented a quantum switch (see Fig. \ref{fig:guo setup}.a) using path as the control for the order and polarisation as the target qubit\cite{Guo_2020}. The Pauli operators are realised using a set of waveplates. The noisy channels are constructed according to the relevant probability distributions (similar to Eq. \ref{channel}) using the individual instantiations of the different Pauli operations.

\emph{Classical communication advantage:}
Similar to the classical communication advantage that was shown by Goswami et al. \cite{goswami2018communicating}, Guo et al. also experimentally demonstrated that two fully depolarising channels acting in an indefinite order can lead to nonzero information capacity \cite{Guo_2020}. They showed this for different input polarisation states $\{\ket{H},\ket{V}\}$, $\{\ket{D},\ket{A}\}$ and $\{\ket{R},\ket{L}\}$, achieving one-shot classical capacities of of $0.0422{\pm}0.0010$ $0.0426 {\pm}0.0011$ $0.0432{\pm}0.0009$ bits, respectively. Fig. \ref{fig:guo setup}.b shows these values compared with the ideal value of 0.049 bits.

\emph{Quantum communication through zero capacity channel:}
The authors used the combination of a bit flip channel $\pazocal{N}_p(\rho){=}(1-p)\rho{+}p \sigma_1 \rho \sigma_1$ and a phase flip channel $\pazocal{N}_q(\rho){=}(1-q)\rho{+}q \sigma_3\rho \sigma_3$ with $p{=}q{=}1/2$. A definite-ordered combination of these channels has zero classical and quantum capacity. However, Salek et al. has proposed that it is possible to transmit heralded, noiseless qubits through the channel $\epsilon _{-}$ (Eq.~\ref{anticommutator_channel} )  if the channels are arranged in an indefinite order\cite{salek2018quantum}. Guo et al. have shown experimentally that indeed this is the case, they achieved a heralded one-shot coherent information ($Q_1$) of $0.812{\pm}0.003$. Fig. \ref{fig:guo setup}.c is a plot of $Q_1$ as a function of the parameter $q$. For comparison,  $Q_1$ of a definite-ordered combination of $\pazocal{N}_p$ and $\pazocal{N}_q$ is shown by the black curve.

\emph{Quantum communication with entanglement-breaking channels: }
Entanglement-breaking channels cannot transmit quantum information \cite{Wilde13}. Two copies of the entanglement-breaking channels $\pazocal{N}(\rho){=}1/2(\sigma_1\rho\sigma_1{+}\sigma_2\rho\sigma_2)$ arranged in a a definite order yields a zero-capacity channel \cite{chiribella2018indefinite}. However, when these channels are in an indefinite order, the resulting auxiliary channels $\epsilon_{\pm}$  (Eq.~\ref{commutator_channel} and ~\ref{anticommutator_channel}) are unitary channels. This results in deterministic noiseless qubit transmission. In the experiment, the authors achieved a one-shot coherent information of $0.855{\pm}0.004$.


\subsection{Communication advantage: further discussion}
Although this is a review of experiments in causality, for the sake of completeness, we are highlighting an active discussion. When the communication enhancement based on superposition of causal order were introduced in Refs. \cite{Ebler_2018,salek2018quantum, chiribella2018indefinite,Goswami_2018}, two alternative resources---coherent superposition of path  by Abbott et al.\cite{abbott2018communication} and coherent superposition of intermediate quantum operations by Gu\'{e}rin et al. \cite{Gu_rin_2019}---were proposed. Both these works showed that the activation---increase from zero to nonzero---of both classical and quantum capacities is possible even for a process that does not exhibit indefinite causal order. Specifically, \citet{abbott2018communication} demonstrated that the superposition of two paths, each containing a depolarising channel can lead to a higher classical capacity than superposition of causal order (with a caveat that the activated capacity is dependent on channel implementation, unlike superposition of causal order). However, it turned out that advantages associated with superposition of path is not universal. For certain combinations of noisy channels---two entanglement breaking channels in the case of quantum communication \cite{chiribella2018indefinite} and one depolarising and one unitary channel in the case of classical communication \cite{Goswami_2018}---superposition of causal order always outperforms superposition of path. The other resource, introduced by Gu\'{e}rin et al.\cite{Gu_rin_2019}, showed that a coherent superposition of intermediate unitary operations (referred to as superposition of direct pure processes, SDPP) can also lead to a communication advantage. As could be expected, superposition of indefinite causal order does not present an advantage over SDPP because of the presence of a clean channel in the latter in the first place.These subsequent works which also show a communication advantage motivates two interesting questions: (1) Can the communication enhancement in the quantum switch be explained by an alternative resource? and (2) How do we put different resources in a comparative framework? 

In response to (1), Gu\'{e}rin et al.\cite{Gu_rin_2019},argues that as noisy operations were acting only on the target qubit, the control qubit provides a clean `side channel' to transmit information.  As a counter to this argument, Kristj\'{a}nsson et al. \cite{kristjnsson2019resource} showed that theoretically there is no side channel generated in the quantum switch and thus indefinite causal order is a resource in itself. One subtle thing to highlight is that in experimental implementations the vacuum plays a part \cite{Oreshkov_2019}.  This is different from the ideal quantum switch which is a mathematical framework dependent solely on the local channels, and the background causal structure. The role of vacuum is prevalent in the protocols of Abbott et al. \cite{abbott2018communication} and Gu\'{e}rin et al. \cite{Gu_rin_2019} (what Kristj\'{a}nsson et al. refers to as \emph{vacuum-extended} channels).

Motivated by (2), Kristj\'{a}nsson et al. \cite{kristjnsson2019resource} presented a resource-theoretic framework of indefinite causal order. A recent experiment by Rubino et al.\cite{Party_paper} compared all three resources---superposition of order, superposition of path, SDPP---and confirmed what has been already been shown in theory that coherent superposition of intermediate unitary operations always yields higher information theoretic advantage compared to the other resources. All these discussions provide a clear motivation for a generalised resource theory to encompass different types of coherent controls  \and the advantages they bring about.

\section{Application in computation and communication complexity}
\begin{figure*}[!t]
\begin{center}
\includegraphics[width=2\columnwidth]{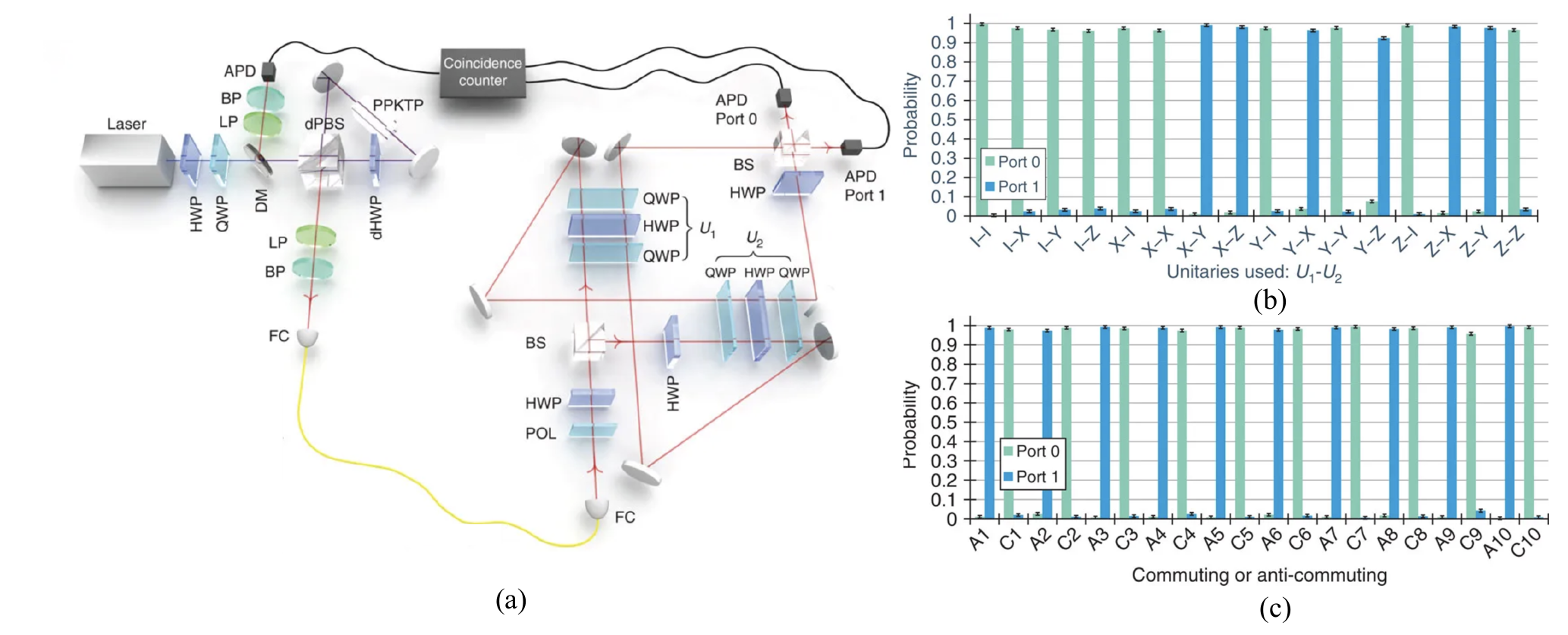}
\vspace{-2mm}
\caption{(a) Experimental setup \cite{Procopio_2015} to distinguish whether a pair of unitary gates commute or anti-commute.  The photons were generated via spontaneous parametric down-conversions. One photon is used as a herald, and the other is the input to the Mach-Zehnder interferometer. The path, which is used as the control, is coherently superposed by the input beamsplitter (BS).  The gates which consist of the train of waveplates in each arm, act on the polarisation (b) The probability that a photon is measured in either Port 0 or Port 1 for different combinations of Pauli gates and identity.  Green (blue) bars correspond to the commuting (anti-commuting) combinations. (c) Same as (b), only that the gates are chosen from a random set of 50 commuting and 50 anti-commuting sets of gates. Reprinted with permission from L.  M.  Procopio \emph{et al.},``Experimental superposition of orders of quantum gates," Nat. Commun. \textbf{6}, (2015) (Ref. \onlinecite{Procopio_2015}). Copyright 2015 by Nat. Commun.}
\label{fig:computational advantage setup}
\end{center}
\end{figure*}
Conventional quantum information theory departed from classical information by considering information carriers that exhibit quantum behaviour. The resulting new paradigm of computation---quantum computation---offers considerable advantage for some specific tasks, e.g. factoring. As we have seen in the previous sections, indefinite causal order adds another aspect of quantum behaviour by allowing for superpositions of the order that operations act on a target system\cite{chiribella09,araujo14}, it is thus natural to ask whether this can result to further computational advantage.    Quantum states in the quantum circuit model evolve according to a sequence of quantum gates that have a fixed order---what advantages arise with coherent control of this order?

In this section, we will focus on experiments which demonstrate that indefinite causal order presents advantages for some computational and communication complexity tasks.  The quantum switch is again central to all these demonstrations.

\subsection{Distinguishing between commuting and anti-commuting gates}

We first focus on the following task: Given that a pair of gates either commute or anti-commute, how do we distinguish between these two cases? A quantum circuit with a fixed order of these gates would require at least two uses of one of the gates. In contrast, a quantum circuit with an indefinite order of these two gates require only a single use of each gate \cite{chiribella12}. 

Consider two gates $U_1$ and $U_2$ acting on a target qubit $\ket{\psi}_t$. A control qubit $\ket{\psi}_c$  controls the order that these two gates are applied: $\ket{0}_c$ leads to the order $U_1U_2$ and $\ket{1}_c$ leads to the order $U_2U_1$. Placing $U_1$ and $U_2$ in a quantum switch and setting the control qubit to a superposition thus results to a superposition in the order that these gates are applied to the target qubit.  More specifically, for an input joint state of $(\ket{0}_c+\ket{1}_c)/\sqrt{2}\otimes\ket{\psi}_t$, and a Hadamard operation applied at the output of the control qubit, the resulting joint state is,

\begin{align}
    \frac{1}{2}\left(\ket{0}_c\otimes\{U_1,U_2\}\ket{\psi}_t+\ket{1}_c\otimes[U_1,U_2]\ket{\psi}_t\right).
\end{align}
If the control qubit is measured to be in $\ket{0}_c$, the gates commute, if the control qubit is measured to be in $\ket{1}_c$, the gates anti-commute.

Procopio et al.\cite{Procopio_2015} showed this advantage using a quantum switch that uses path as control and polarisation as the target qubit for the gates $U_1$ and $U_2$. They used a quantum switch similar to the one used by Rubino et al.\cite{rubino2017} for testing the causal witness, in this case replacing $M^A$ and $M^B$ by $U_1$ and $U_2$. Photons from spontaneous parametric down-conversion were used, one as a herald and one as input to the quantum switch. As shown in Fig. \ref{fig:computational advantage setup}.a, the quantum switch consists of a Mach-Zender interferometer with each arm containing  a loop. The train of quarter-waveplates and half-waveplates in each arm act as the gates. The input 50/50 beamsplitter creates the superposition of  transmitted ($\ket{0}_c$) and reflected ($\ket{1}_c$) paths for the control qubit.   The two paths are again coherently combined at the final 50/50 beamsplitter, which acts as the Hadamard operation for the control qubit. The direction that the photon exits---measured by the single-photon detectors at either port 0 or port 1---reveals whether the gates are commuting or anti-commuting.

Fig. \ref{fig:computational advantage setup}.b. shows the probability that a photon exits from a port, given that the quantum switch has $U_1$ and $U_2$ as the gates. Note that $U_1$ and $U_2$ were chosen from the three Pauli operations $\{\sigma_x, \sigma_y, \sigma_z\}$ and the identity operation $\mathbb{1}$.  Fig. \ref{fig:computational advantage setup}.b shows the results for all 16 combination of gates. As expected, when the gates are commuting, the photon exits port 0 (green bars), otherwise if the gates are anti-commuting, the photon exits port 1 (blue bars).  The success rate---defined as the probability to exit the correct port---is $0.973 \pm 0.016$.  Fig. \ref{fig:computational advantage setup}.c shows the results for 50 randomly generated commuting unitaries and 50 randomly generated anti-commuting unitaries.  The authors report a success of $0.976 \pm 0.015$ in this case. Note that these success rates are both larger than the highest success rate possible with a definite-order quantum circuit which is $0.9288$.

\begin{figure*}[!t]
\begin{center}
\includegraphics[width=2\columnwidth]{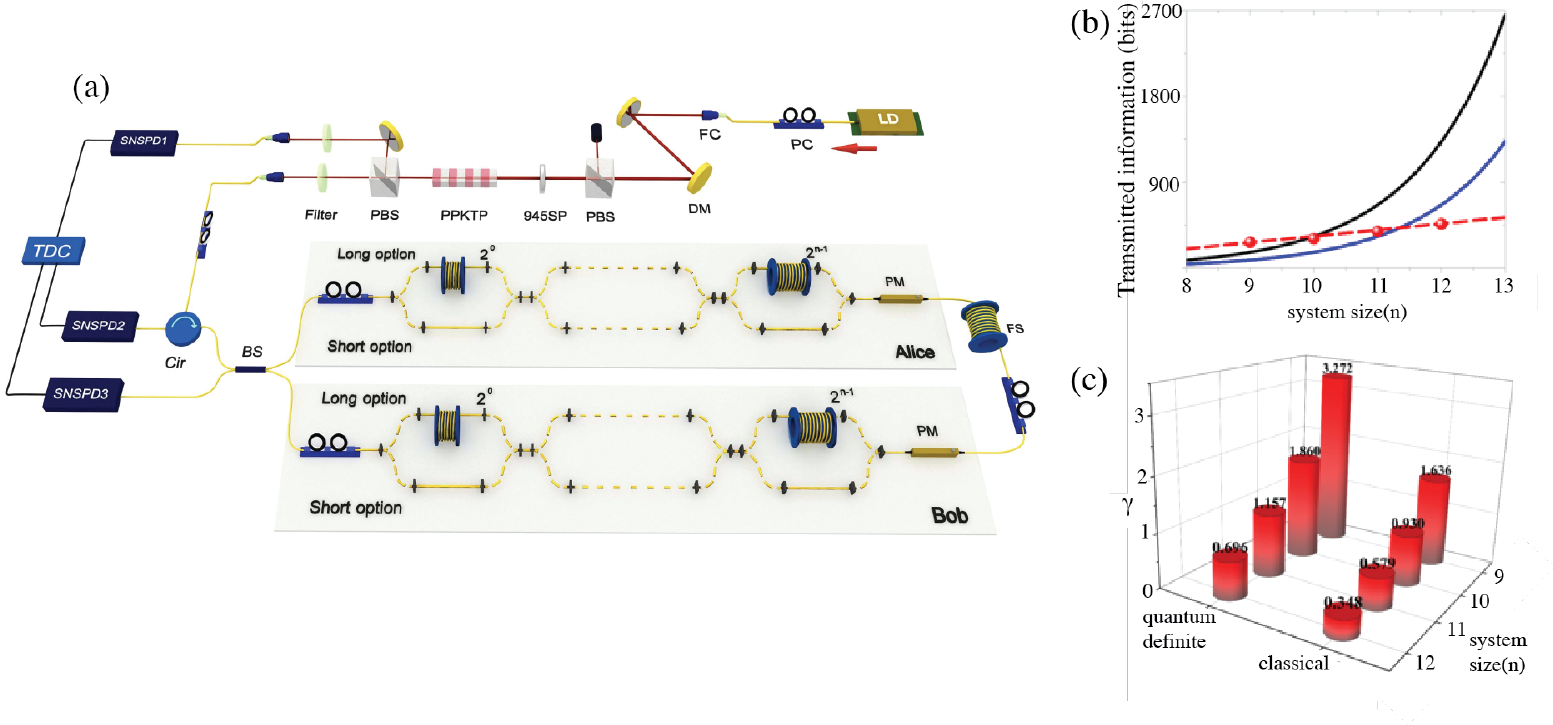}
\vspace{-2mm}
\caption{(a) Experimental setup for showing an advantage in communication complexity \cite{Wei_2019}. Photon pairs were produced from spontaneous parametric down-conversion. One photon is used as a herald and the other is fed to the interferometer via circulator (Cir) and beamsplitter (BS) which prepares the control in a superposition of the two paths. In the Sagnac loop, Alice and Bob implement the unitary operations by using variable delays and a phase modulator (PM). The binary digits were represented by either the long option or the short option. For the long option, the authors stored the pulse train by using a 7 km long fiber spool (FS). The result of the computation was obtained from the outputs of the superconducting nanowire single photon detectors (SNSPD2 and SNSPD3). (b) The transmitted information with respect to the system size($n$).  Red points show the experimental results with the best fit line using a quantum switch. For $n {=} 12$  both classical (back) and quantum causally definite (blue) protocols require more transmitted bits, indicating a clear advantage for using a quantum switch. (c) The parameter $\gamma$ represents the ratio of the transmitted information $Q$ over the lower bound on the transmitted information $C$ for a given quantum or classical causally separable protocol.  For $n=12$, $\gamma<1$ indicating that a quantum switch requires less transmitted bits for the EE game. Reprinted with permission from K. Wei \emph{et al.}, ``Experimental quantum switching for exponentially  superior  quantum  communication  complexity," Phys. Rev. Lett. \textbf{122}, (2019) (Ref. \onlinecite{Wei_2019}). Copyright 2019 by APS.}
\label{fig:communication_complexity_setup}
\end{center}
\end{figure*}

\subsection{Communication complexity}

Communication complexity quantifies the cost of communication requirements in solving problems when the input is distributed to two or more parties \cite{chuang00}. \citet{Guerin2016} suggested that indefinite causal order can lead to a reduction in the communication complexity of the \emph{exchange evaluation (EE)} game. Consider bit strings $\vec{x}$, $\vec{y} \in \{0, 1\}^n$ and Boolean functions $f, g : \{0, 1\}^n \rightarrow \{0, 1\}$ with $f(\vec{0}) {=} g(\vec{0}) {=} 0$, where $\vec{0}$ is a string of zeros. Alice has $(\vec{x}, f)$ and Bob has $(\vec{y}, g)$, and  both of them can signal a third party, Charlie. Charlie's task is to evaluate the \emph{exchange evaluation function} $\mathrm{EE}(\vec{x}, f, \vec{y}, g) {=} f(\vec{y}) {\oplus} g(\vec{x})$, where $\oplus$ denote addition modulo 2. The communication is strictly one-way i.e. Charlie cannot signal to Alice and Bob. 

\citet{Guerin2016} proposed that superposition of causal order between Alice and Bob---such that communication direction is also in a superposition---can achieve deterministic, exponential communication advantage in terms of the number of bits that Alice and Bob communicate. Charlie can compute the exchange evaluation function by measuring the control qubit of a quantum switch. 

\citet{Wei_2019} implemented a variant of the protocol proposed by \citet{Guerin2016} using a quantum switch that uses path as a control for the order. For the target system, they used $ d {=} 2^{n+1}$-qudit obtained from the arrival time of the same photon.  The arrival time was divided into  $2^{n+1}$ time bins which then make up an orthonormal basis $\ket{z}$ with $z{\in} \{0, 1, \dots, 2^{n+1}−1\}$.  The $EE$ game can be translated to a game that distinguishes commuting unitaries from anti-commuting unitaries, similar to the task in the previous section. \citet{Wei_2019} introduced two unitaries $U_A \equiv X^x_d D_d(f)$ and $U_B \equiv X^y_dD_d(g)$, where $D_d(f)=Z_d^{f(z)d/{2z}}$ and $X_d$ and $Z_d$ are high-dimensional Pauli operators acting on the basis-set $\{\ket{z}\}$ which satisfy $Z_dX_d=\exp(2\pi i/d)X_d Z_d$. With these definitions, $[U_A, U_B] \ket{0}_t = 0$ if $f(y) \oplus g(x) = 0$, $\{U_A, U_B\} \ket{0}_t = 0$ if $f(y) \oplus g(x) = 1$, where the subscript $t$ denotes the target system. 

The problem can be solved by using a quantum switch with operations $U_A$ and $U_B$, and input control in $\ket{+}_c = 1/\sqrt{2} (\ket{0}_c + \ket{1}_c)$, and  input target in $\ket{0}_t$. A Hadamard operation applied to the control qubit results to an overall state 
\begin{align}
\frac{1}{2}\left(\ket{0}_c\otimes\{U_A,U_B\}\ket{0}_t{-}\ket{1}_c\otimes[U_A,U_B]\ket{0}_t\right).
\end{align}
Measuring the control qubit in the basis $\ket{z}$ then reveals whether $f(\vec{y}) {\oplus} g(\vec{x})=0$ or $f(\vec{y}) {\oplus} g(\vec{x})=1$.

The experiment uses a fibre-based Sagnac interferometer, as shown in Fig.~\ref{fig:communication_complexity_setup}.a\cite{Wei_2019}. Photon pairs are produced from SPDC, one of the photons is used as a herald and another is fed to the interferometer.  The preparation of the control qubit into a superposition of paths is done by a combination a circulator and a fibre beamsplitter. Alice and Bob in the Sagnac loop each have variable delays and a phase modulator to implement the  unitary operations $U_A$ and $U_B$. Alice and Bob individually uses the phase modulator twice, once when the photon enters the Sagnac loop, and once when it exits the loop. The photon interference in the loop ensures that it exits at either of the output ports of the interferometer depending on the answer to the EE game.

Fig.~\ref{fig:communication_complexity_setup}.b  shows the transmitted bits as a function of system size $n$ (i.e. for when there are $2^n$ choices of delay for Alice and Bob). The red dashed line is the best fit for the experimental data. For comparison, the black and blue curves show the required transmitted bits for the best classical protocol and the best fixed-order quantum protocol, respectively. For a system size $n=12$, the indefinite-ordered protocol using the quantum switch beats both classical and best fixed-order quantum protocols. 
Fig.~\ref{fig:communication_complexity_setup}.c compares the performance of the indefinite-ordered protocol to fixed-order processes by taking the ratio of transmitted information through the quantum switch ($Q$) and the minimum transmitted information through either causally separable quantum or classical protocol $C$, i.e. $\gamma=Q/C$.   For $n=12$, this ratio is less than for both the classical and quantum-definite ordered processes indicating a clear advantage in using a quantum switch.

\begin{figure*}[!t]
\begin{center}
\includegraphics[width=1.7\columnwidth]{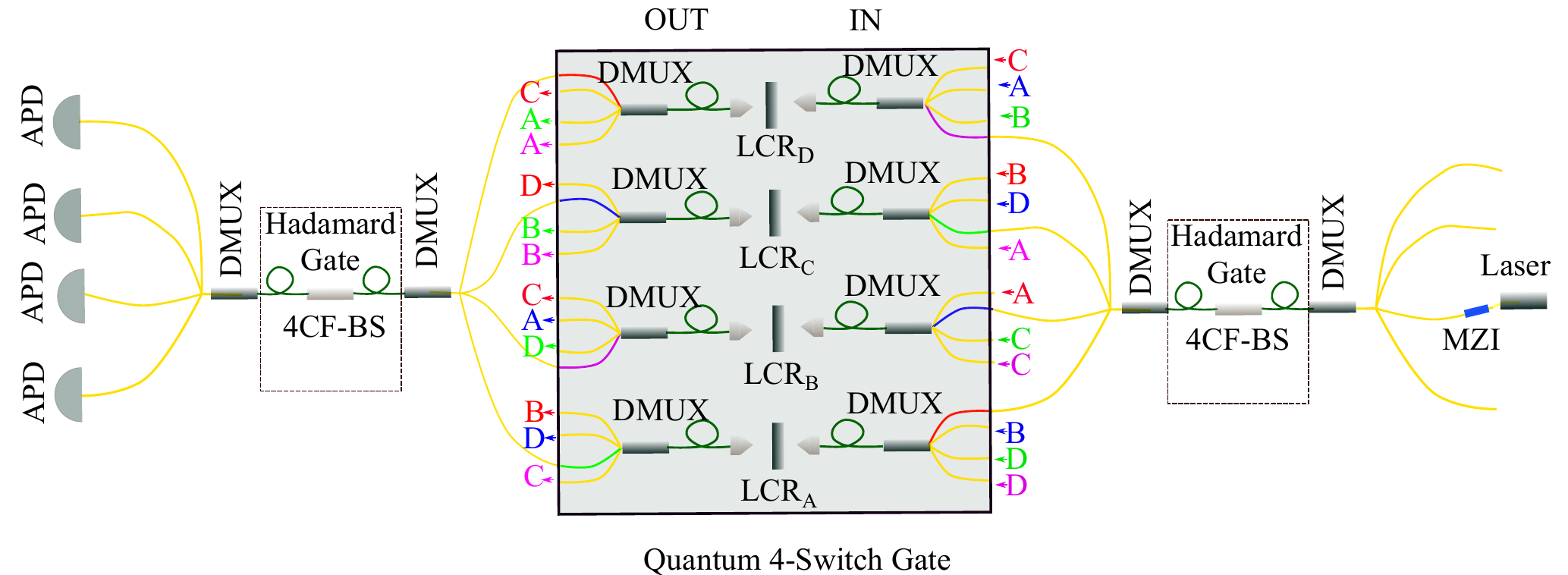}
\vspace{-2mm}
\caption{Experimental setup for the quantum 4-switch, as described in Ref. \onlinecite{taddei2020experimental}. An input photon was coherently divided into four spatial modes using a four-core fibre beamsplitter (4CF-BS), placed between two multiplexer/demultiplexer (DMUX) units. The four output modes were then sent to the quantum 4-switch gate. Each spatial mode was associated with a unique permutation of the four unitary
polarisation operations. The photons enters through the `IN' side and exit through the `OUT' side as shown. The notation, for example, ``$\leftarrow${A}" means ``from A" and ``A $\leftarrow$" means ``to A". After leaving the quantum-4 switch, the four spatial modes were recombined using a second 4CF-BS. The outputs were connected directly to single-photon detectors (APD).}
\label{fig:quantum-4-switch}
\end{center}
\end{figure*}

\subsection{Hadamard promise problem}

So far we have only considered a quantum switch where the control is a qubit---controlling the order of only two operations. The quantum switch can be generalised to a quantum $n$-switch, with a control that specifies the order of $n$ operations. In this case, a $d$-dimensional control system  (where $d\leq n!$) creates a coherent superposition of $d$ permutations of all the quantum operations.\cite{araujo14}  To formulate the generalisation, consider a set of unitary operations $\{U_i\}$. We define $\Pi_x{=}U_{\sigma_x(n)}.U_{\sigma_x(n-1)}{\dots}U_{\sigma_x(1)}$ with $\sigma_{x(i)}$ being the $i$-th element of the $x$-th permutation of the gates, where $x{\in}\{1,2,\dots,d\}$. The action of the quantum switch $T_n$ on the input state $\ket{x}_c\otimes\ket{\psi}_t  $ becomes,

\begin{align}
    T_n( \ket{x}_c\otimes \ket{\psi}_t){=}\ket{x}_c \otimes\Pi_x \ket{\psi}_t .
\end{align}
Taddei et al.\cite{taddei2020experimental} used this generalisation of a quantum switch to show a computational advantage for the Hadamard promise problem: Given a $d$-dimensional Hadamard matrix $M_d$\cite{hadamard_matrix_1978} with  all $+1$ entries  along  its  first  row  and  first column,  and a unitary-gate oracle $\pazocal{O}$ fulfilling the promise $\Pi_x{=}m_{x,y}\Pi_0$---where $\{m_{x,y}\}={\pm}1$ are the entries of $M_d$---for some column $y\in[P]$ of $M_d$, compute $y$. Note that here, $[P]$ is defined as $[P]:={0,1,...P-1}$ ,i.e. $P$ permutations of an $n$-letter alphabet. 

After the series of steps: (1) initialise joint system to $\ket{0}_c\otimes\ket{\psi}_t$ (2) apply $M_d$ on the control (3) apply $T_n$ on the joint system (4) apply $M_d^{-1}$ on the control, the resulting state is,
\begin{align}
    \frac{1}{d} M_d^{-1}T_n M_d (\ket{0}_c \otimes \ket{\psi}_t  ){=}\ket{y}_c \otimes \Pi_0 \ket{\psi}_t.
\end{align}
From this state, the value of $y$ can be read by measuring the control in the computational basis.

Using the quantum $n$-switch, the total number of oracle queries---the query complexity $R$---of the Hadamard promise problem is $R{=}n$ for all $P{\leq} n!$.   At least for the case of $P{=}n!$, this is quadratically lower in $n$ when compared with the best definite-ordered algorithm.

Fig.~\ref{fig:quantum-4-switch} shows the experimental setup used to verify this computational advantage\cite{taddei2020experimental} for the case of $d{=}4$, i.e. using a quantum 4-switch. The path was used to control the different permutations of the order of the operations. The implementation is based on multi-core optical fibres where it is convenient to separate the photon into different paths. Polarisation (of the same photon) was used as the target qubit.  In Fig.~\ref{fig:quantum-4-switch} 1546 nm pulses are sent through a combination of demultiplexers (DMUX) and 4-core fibre beamsplitter (4CF-BS) to realise the Hadamard operation.
 
Each single-mode output in this operation becomes an input to the quantum 4-switch, where each one corresponds to a different permutation of the polarisation operations ${U_i}$.   Each single-mode fibre input of the quantum 4-switch connects to a different 4-core fibre unit.  The different permutations of the polarisation operations $U_i$ were coherently applied and realized using controllable liquid crystal retarders (LCR). After the LCR light is coupled back into another 4-core fibre on the OUT-side, which are in turn connected via a demultiplexer to single-mode fibres that connect to the next 4-core fibre back on the quantum 4-switch's IN-side. After the quantum 4-switch, a second Hadamard operations is applied to the control using another DMUX-4CF-BS-DMUX train. The output of this is directly connected to single-photon detectors, where the detection of a photon in $y$-th detector gives the value of $y$ in the Hadamard promise problem.

For two sets of unitary operations\cite{taddei2020experimental}, one consisting of $\{\mathbb{1}, Z, X\}$ and one consisting of $\{\mathbb{1},Z,X, (Z{+}X)/\sqrt{2}\}$ the average probability of success were $0.948{\pm}0.005$ and $0.959{\pm}0.008$ respectively. This success rate was achieved using only 4 queries, compared to 9 queries for the best known fixed-order quantum circuit that can solve the Hadamard promise problem.

\section{Summary and Concluding Remarks}

We have presented the experiments in the nascent field of quantum causality. To facilitate the logical flow of this review, the experiments were not presented in chronological order. The first experiment was on the computational advantage in distinguishing commuting and anti-commuting gates \cite{Procopio_2015}, followed by works on detecting indefinite causal order \cite{rubino2017, rubino2017b, Goswami_2018}, works that show communication complexity and communication capacity advantages \cite{Guo_2020, goswami2018communicating, Wei_2019}, and most recently the quantum 4-switch \cite{taddei2020experimental}. The field is enlivened by both foundational and practical aspects. Indefinite causal order will most likely likely be a feature of any theory that successfully combines general relativity and quantum mechanics and we enthusiastically await experiments that will have foundational importance in this area. In the interim, we have provided a summary of the equally important first experiments in quantum causality which are mostly pragmatic in nature.

As shown by the experiments of \citet{rubino2017} and \citet{Goswami_2018}, a causal witness enables us to detect indefinite causal order in a process like the quantum switch. More strongly, \citet{rubino2017b} have shown that indefinite causal order can be witnessed in a theory-independent manner using a Bell's theorem formulated for temporal order. The quantum switch has been used to enable communication through depolarising channels---which normally scrambles any information whether classical or quantum---as demonstrated in the experiments of \citet{goswami2018communicating} and \citet{Guo_2020}.  Computational advantage has also been demonstrated for tasks that can be considered as variants of channel discrimination. \citet{Procopio_2015} have shown that indefinite causal order in a quantum switch can be used to distinguish beteween commuting and anti-commuting gates with only one use of each gate. A reduction in communication complexity was shown by \citet{Wei_2019} by casting the exchange evaluation game as a task of distinguishing between commuting and antic-commuting operations. \citet{taddei2020experimental} demonstrated that using a quantum switch to solve the Hadamard promise problem leads to a quadratically smaller number of queries compared with the best definite-ordered algorithm. These practical demonstrations reinforce the idea that indefinite causal order is a resource which can be exploited for various communication and computation tasks.

Although all the experiments we have reviewed are photonic in nature, there is no fundamental reason that indefinite causal order cannot be achieved in other platforms.  Photonic implementations of the quantum switch abound because it is convenient, a photon is an indivisible system that possesses several degrees of freedom which can be easily controlled and measured. Any two degrees of freedom can serve as control and target systems for the quantum switch. Current implementations of the quantum switch can be grouped broadly in terms of the degree of freedom used as control: path or polarisation.
Using path as the control means that the various input and output locations within the same optical element---e.g. where the photons impinge on a waveplate as in Fig. \ref{fig:computational advantage setup}.a---are not co-located.  These implementations assume that regardless of the path that a photon takes, the same set of operations act on the target system. We emphasise that this is a reasonable assumption. We mention this only to clarify any confusion that might arise when one compares the theoretical proposal of a quantum switch---where the maps themselves have one input and output port---with an experimental implementation.  If one uses polarisation as a control for the order, it is possible for the operations to be spatially co-located regardless of the order, hence operations are not spatially distinguishable. However, this leaves other degrees of freedom to be used as the target system, e.g. transverse spatial mode or time-bins, which are practically more challenging to manipulate compared to polarisation. 

Most of the experiments so far (with the exception of \citet{taddei2020experimental}) used a qubit for the control.  This is limited to indefinite causal order  between only two operations. Whereas this has not stopped experiments and theorists alike from showing the advantages to be had from having indefinite causal order, it is of interest whether higher-dimensional extensions can lead to better advantages, and whether these can also be shown experimentally. Having qudits both for control---to enable more permutations of the order---and target---to increase information capacity---can lead to fruitful extensions especially towards scalability. Multi-core fibre optic technology has already enabled the first  quantum 4-switch, we look forward to further experiments that involve qudits using maturing technologies that manipulate all the photonic degrees of freedom \cite{mounaix2019time}. 

As a final remark, we point out that although our experiments clearly reproduce the output of a quantum switch, there remains the subtlety that all our implementations are a function of the channels in the quantum switch and the vacuum. For a proper accounting in a resource theory of indefinite causal order, we specify that if the input to a channel is the vacuum, the channel acts as an identity channel\cite{kristjnsson2019resource}.  Although this does not stop experiments from showing the various communication or computational advantages that have been or will be discovered, this subtlety will certainly play a role in evaluating various resource theories of coherent control.

\begin{acknowledgments}
We thank the editors for the invitation to write this review. We thank Dr. Fabio Costa, Dr. Cyril Branciard, Dr. M\'{a}rcio Taddei, Prof. Giulio Chiribella, and anonymous referees for useful discussions. This work has been supported by: the Australian Research Council (ARC) by Centre of Excellence for Engineered Quantum Systems (EQUS, CE170100009),  the RTP scholarship from the University of Queensland for KG,  and Westpac Research Fellowship and L'Oreal-UNESCO FWIS grant for JR.  We thank Dr Markus Rambach and Michael Kewming for reading early drafts of this review. We acknowledge the traditional owners of the land on which the University of Queensland is situated, the Turrbal and Jagera people.  
\end{acknowledgments}

\section*{Data availability statement}
The data that supports the findings of the articles reviewed in this review paper are available within the articles and their supplementary material.
\appendix

\bibliography{AVS.bib}

\end{document}